\documentclass[twoside,english]{elsarticle}
\usepackage{graphicx,amsmath,amsfonts,amssymb,lmodern,multirow}
\usepackage[utf8]{inputenc}
\usepackage[T1]{fontenc}

\makeatletter
\journal{Physica B}

\makeatother

\providecommand{\theoremname}{Theorem}

\begin{document}
\title{Limiting effects of geometrical and optical nonlinearities on the
squeezing in optomechanics}

\author[Lab1]{P. Djorwé\corref{cor2}}

\address[Lab1]{Laboratory of Modelling and Simulation in Engineering,
Biomimetics and Prototypes, Faculty of Science, University of Yaoundé I,
Cameroon}

\author[Lab2]{S.G. Nana Engo\corref{cor1}}

\ead{snana@univ-ndere.cm}

\address[Lab2]{Laboratory of Photonics, Faculty of Science, University of
Ngaoundéré, Cameroon}

\author[Lab1]{J.H. Talla Mbé}

\author[Lab1]{P. Woafo}

\cortext[cor1]{Corresponding author}

\cortext[cor2]{Principal corresponding author}

\begin{abstract}
In recent experiments, the re-thermalization time of the mechanical resonator is
stated as the limiting factor for quantum applications of optomechanical
systems. To explain the origin of this limitation, an analytical nonlinear
investigation supported by the recent successful experimental laser cooling
parameters is carried out in this work. To this end, the effects of geometrical
and the optical nonlinearities on the squeezing are studied and are in a good
agreement with the experimental results. It appears that highly squeezed
state are generated where these nonlinearities are minimized and that high
nonlinearities are limiting factors to reach the quantum ground state.
\end{abstract}

\begin{keyword}
Quantum optomechanics \sep nonlinear effects  \sep squeezing
\PACS 42.50.Wk \sep 42.50.Lc \sep 42.79.Gn \sep 37.10.De
%  \MSC[2008]23-557
\end{keyword}
\maketitle

\section{Introduction}

Squeezing is a beautiful quantum phenomenon with amazing potential applications
\cite{c1,c2} of which the most recent are connected with continuous variables
quantum information \cite{c3,c4} and ultrasensitive measurement of weak
perturbations as the gravitational waves \cite{c5,c6,c7}. Squeezed states are
nonclassical states in which the variance of at least one of the canonical
variables is reduced below the noise level of zero point fluctuations. To
generate squeezed states, the common technique consists to use an optical cavity
filled with a nonlinear Kerr medium which is fed with an external pumping field
\cite{c8}. With the recent advances in cooling techniques for nano scale
optomechanical systems, various setups have been designed for quantum ground
state engineering of mechanical mirrors with highly squeezed states of light
\cite{c9,c10,c11,c12,c13,c14}. Indeed, with such technique it is now possible to
obtain effective phonon number less than $1$ \cite{c16,c17,c18,c19,c20,c21}. The
limiting factors to obtain much lower phonon number are the re-thermalization
time of the mechanical resonator $\tau_{th}=\frac{\hbar Q_m} {k_{B}T}$ (where
$\hbar$ is Planck's constant, $Q_m$ is the mechanical quality factor, $k_B$ the
Boltzmann constant and $T$ the temperature of the support), which competes with
the cooling, and the ubiquitous phase noise of the input laser which can create
a discrepancy between experimental results and theoretical prediction
\cite{c16,c17}. Nevertheless, a lot of theoretical studies on the subject has
been carried out in the last decade and several proposals have been produced
\cite{c151,c152,c153}. In Ref. \cite{c153} we applied the technique of
back-action cooling to show that the cooling of the nanomechanical oscillator to
its ground state is limited by the effects of both optical and mechanical
nonlinearities.

In this paper, by using the parameters of the experimental laser cooling
of Ref. \cite{c16}, we extend the previous treatment to show through analytical
study that there are the nonlinearities which limit the squeezing in
optomechanics. The first
one which depends on the geometry of the mechanical structure is known as the
geometrical nonlinearity derives from the nonlinear dynamics of the beams
\cite{c154,c22,c23}. The second one is the optical nonlinearity which appears as
a
nonlinear phase shift \cite{c10}. The geometrical nonlinearity which is always
present and not negligible in nano resonators, is shown to be a limiting factor
to reach the quantum ground state as suggested in Ref. \cite{c23}. In the same
way, it is shown that high absolute values of the optical nonlinearity limit the
squeezing of the output intensity.

The paper is structured as follows. Section \ref{sec:Dyn} will set the stage for 
exploring the dynamics of our system, deriving in particular the nonlinear 
Quantum Langevin Equations and the linearized equations of motion. Sections 
\ref{sec:SqMec} and \ref{sec:SqOpt} subsequently makes use of numerical 
simulations to discuss the squeezing of the mechanical and the optical output 
quadratures. Finally, we conclude with an outlook of possible future directions.

 \section{Dynamics equations}\label{sec:Dyn}

We consider an optomechanical resonator described by Fig. 1 of Ref. 
\cite{c16}. The dynamics equations of a mechanical oscillator coupled to a 
driven cavity is usually derived from a single-mode Hamiltonian 
\cite{c12,c153,c154,c25} as,
\begin{subequations}
\label{eq:01}
\begin{align}
&\ddot{x}_m +\Gamma_m\dot{x}_m+\Omega_m^2x_m - \beta''\Omega_m 
=g_M\Omega_m|\alpha(t)|^2+\frac{F_{th}}{Mx_{ZPF}}, \\
& \dot{\alpha} =\left[i(\Delta +\frac{g_M}{x_{ZPF}}x)-\frac{\kappa}{2}\right] 
\alpha(t)-i\varepsilon^{in}+\sqrt{\kappa}\alpha^{in},
\end{align}
\end{subequations}
$x_m$ and $p_m$ are the dimensionless position and momentum operators of the
mechanical oscillator related to their counterparts operators of the
nanobeam as $x=\sqrt{\frac{\hbar }{2M\Omega_m}}x_m=x_{ZPF}x_m$ and
$p=\frac{\hbar}{x_{ZPF}}p_m$, with $[x_m,p_m]=2i$. The parameters
$g_M=\sqrt{2}\omega_c\frac{x_{ZPF}}{d_{0}}$, $d_{0}$, $\Omega_m$ and $
\varepsilon^{in}$ are respectively the optomechanical coupling, the cavity
length, the mechanical frequency and the amplitude of the input laser beam. The
cavity decay rate and the mechanical damping of the mechanical oscillator are
respectively represented by $\kappa $ and $\Gamma_m$. The laser-cavity detuning
is $\Delta =\omega_{\ell}-\omega _c$ with $\omega_c$ the optical cavity mode
frequency and $\omega_\ell$ the laser frequency. The terms $F_{th}$ and
$\alpha^{in}$ represent the Langevin force fluctuations and the input laser
fluctuations. The terms $\beta''=\frac{\beta'x_{ZPF}^2x_m^3}{\Omega_m}$ and
$g_Mx_m\alpha$ represent the mechanical and optical anharmonic terms. When the 
nanoresonator is subjected to a large displacement amplitudes, it displays a 
striking nonlinearity $\beta''$ in its response. This comes about because the 
flexure causes the beam to lengthen, which at large amplitudes adds a 
significant correction to the overall elastic response of the beam \cite{c23}. 
The optical anharmonic term is another kind of Kerr medium, which has a 
mechanical origin: the radiation pressure induces a coupling between the 
position of the doubly-clamped flexural resonator and the phase-intensity of the 
light beam, thus modifying the optical path known as the phase shift.

One can derived from the set of equations Eqs.(\ref{eq:01})the following 
nonlinear Quantum Langevin Equations (QLEs) \cite{c153}
\begin{subequations}
\label{eq:1}
\begin{align}
\dot{x}_m &=\Omega_mp_m\\
\dot{p}_m &=-\Omega_mx_m-\Gamma_mp_m+g_M\alpha^{\dag}\alpha + \beta''+F_{th} \\
\dot{\alpha} &=\left[i(\Delta +g_Mx_m)-\frac{\kappa}{2}\right]\alpha
-i\varepsilon^{in}+\sqrt{\kappa}\alpha^{in}\label{eq:a1}\\
\dot{\alpha}^{\dag} &=\left[-i(\Delta +g_Mx_m)-\frac{\kappa}{2}\right]
\alpha^{\dag}+i\varepsilon^*+\sqrt{\kappa}\alpha^{in\dag}.
\end{align}
\end{subequations}

By setting the time derivatives to zero in the set of nonlinear Eqs. 
(\ref{eq:1}), the stationary values of the position of the oscillator and the 
amplitude of the cavity field are
\begin{equation}
\bar{x}_m=2\frac{g_M}{\Omega_m}|\bar{\alpha}|^2,\;\;|\bar{\alpha}|^2=\frac{2
\kappa P_{in}}{\hbar\omega_{\ell}\left((\Delta+g_Mx_m)^2+\frac{\kappa^2}{4}
\right)}.
\label{eq:bb}
\end{equation}%
The values of $\bar{x}$ obey the following third order algebraic equation,
\begin{equation}
\bar{x}^3+\frac{2\Delta x_{ZPF}}{g_M}\bar{x}^2+(4\Delta^2+\kappa^2)\frac{
x_{ZPF}^2}{4g_M^2}\bar{x}-\frac{4\kappa x_{ZPF}^3P_{in}}{\hbar\Omega_m\omega
_{\ell}g_M}=0.
\label{eq:b}
\end{equation}
From Eqs. (\ref{eq:bb}) and (\ref{eq:b}), it appears that both
$\bar{\alpha}$ and $\bar{x}_m$ increase when the input laser power $P_{in}$
increases.

Using the experimental parameters of Ref. \cite{c16} at the detuning of $\Delta
=\Omega_m$ and for $P_{in}=1\operatorname{mW}$, we obtain the following values
of $\bar{x}$ which are in the range of those obtained experimentally in Refs.
\cite{c14,c18}: $1.28\times 10^{-13}$, $-1.09\times10^{-8}+7.43\times10^{-10}i$,
$-1.09\times 10^{-8}-7.43\times 10^{-10}i$. The first solution, which is real
and small, corresponds to the stable regime of the mechanical resonator, while
the two conjugate others, which have the same module ($|\bar{x}|\approx1.09
\times 10^{-8}$), correspond to the unstable regime.

For $|\bar{\alpha}|\gg 1$ (satisfied in Ref. \cite{c16}), the above QLEs can be
linearized by expanding the operators around their steady states: $x_m=\bar{x}_m
+\delta x_m$ and $\alpha =\bar{\alpha}+\delta\alpha$. By introducing the vector
of quadrature fluctuations $u(t)= (\delta x_m(t), \delta p_m(t), \delta I(t),
\delta\varphi(t))^T$ and the vector of noises $n(t)= (0,F_{th}(t),\sqrt{\kappa}
\delta I^{in}(t),\sqrt{\kappa}\delta\varphi^{in}(t))^T$, where $\delta
I=(\delta\alpha^{\dag}+\delta\alpha)$, $\delta\varphi=i(\delta\alpha^{\dag}
-\delta\alpha)$ are the intracavity quadratures of the intensity and the phase ,
and the corresponding hermitian input noise operators $\delta I^{in}$ ,
$\delta\varphi^{in}$, the linearized dynamics of the system can be written in a
compact form
\begin{subequations}
\label{eq:4}
\begin{equation}
 \dot{u}(t)=Au(t)+n(t),
\end{equation}
with
\begin{equation}
 A=\begin{pmatrix}0 & \Omega_m & 0 &0\\\Omega_m(\beta-1) & -\Gamma_m &
G & 0\\0 & 0 & -\frac{\kappa}{2} & -\tilde{\Delta}\\ G & 0 & \tilde{\Delta} &
-\frac{\kappa}{2}\end{pmatrix},
\end{equation}
\end{subequations}
The higher order of fluctuations are safely neglected. The linearized QLEs show
that the mechanical mode is coupled to the cavity mode quadrature fluctuations
by the effective optomechanical coupling $G=g_M|\bar{\alpha}|$, which can be
made large by increasing the input laser power $P_{in}$. $\beta=\frac{3\beta'
x_{ZPF}^2 \bar{x}_m^2}{\Omega_m^2}$ and $\tilde{\Delta} =\Delta+g_M\bar{x}_m$
denote the dimensionless geometrical nonlinearity and the effective detuning.
The range values of the geometrical and optical nonlinearities
are given in the Table \ref{tab:aa}. One remarks that $\beta$ and $\eta$
increase when $\bar{x}_m$ increases and they reach their maximum values at the
detuning $\Delta\approx\Omega_m$.  As expected in the Table \ref{tab:aa}, the
optical and the mechanical
effects are respectively highly pronounced at the optical ($\Delta\approx 0$)
and the mechanical ($\Delta \approx\Omega_m$) resonances \cite{c12}. This leads
us to investigate the squeezing at this particular sidebands.

% \begingroup
% \squeezetable
\begin{table}[ht]
\centering %
\begin{tabular}{c|p{3.5cm}|p{5cm}}
\hline\hline
Detuning $\Delta$ & Mean displacement of the nanobeam $\bar{x}
(\operatorname{m})$ & Range of values of nonlinearities\\\hline
\multirow{2}{*}{$0$ } & $2.77\times10^{-11}$  & $\eta\in[2.54\times10^{-3};
6.79\times10^{-2}]$ \\\cline{2-2}
 & $7.42\times10^{-10}$ & $\beta\in[7.87\times10^{-6};5.72\times10^{-4}]$\\
\hline
\multirow{2}{*}{$\Omega_m$ } & $1.27\times10^{-13}$  &
$\eta\in[1.17\times10^{-5};1]$\\\cline{2-2}
 & $1.09\times10^{-8}$  & $\beta\in[1.66\times10^{-10};1.22]$\\\hline\hline
\end{tabular}
\caption{The range of values of the optical nonlinearity $\eta$ and
the geometrical nonlinearity $\beta$ at the detuning $\Delta=0$ and
$\Delta=\Omega_m$ respectively, using the parameters of Ref.\protect
\cite{c16}.}
\label{tab:aa}
\end{table}
% \endgroup

\section{Squeezing of the mechanical quadratures}\label{sec:SqMec}

\begin{figure}[htpb]
\centering
 \includegraphics[scale=.70]{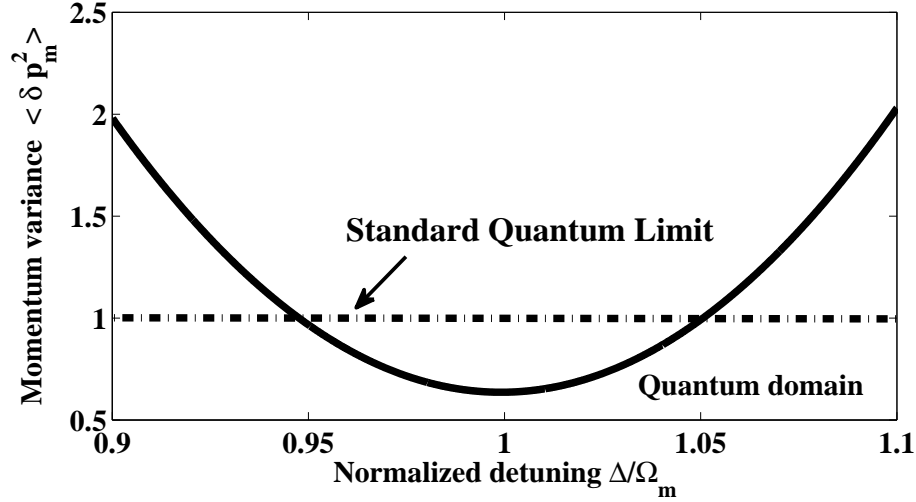}
\caption{Plot of the momentum variance $\langle\delta p_m^2\rangle $ versus 
normalized detuning $\frac{\Delta}{\Omega_m}$ for ${\eta = 0}$, using 
experimental parameters of Ref.\protect \cite{c16}. The value of 
$\frac{\Delta}{\Omega_m}= 1$ corresponds to the momentum variance 
$\langle\delta p_m^2\rangle = 0.6362 $ which means that the momentum is squeezed 
up to about $37\%$.}
\label{fig:Fig1}
\end{figure}

\begin{figure}[htpb]
\centering
\includegraphics[scale=.70]{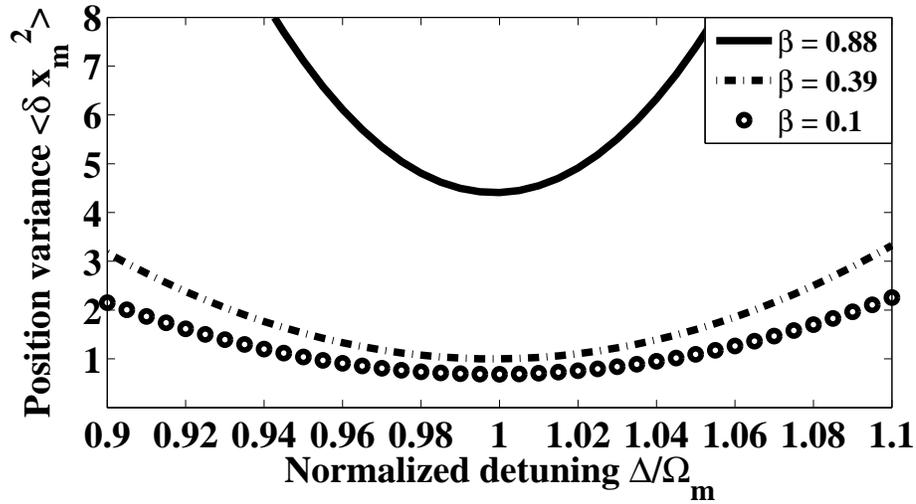}
\caption{Plot of the position variance $\delta x_m^2\rangle$ versus normalized 
detuning $\frac{\Delta}{\Omega_m}$ for different values of $\beta$ with $\eta = 
0$. The dot dashed line is plotted for $\beta =0.39$ and corresponds to the 
Standard Quantum Limit ($\frac{\Delta}{\Omega_m} = 1$; $\langle\delta 
x_m^2\rangle = 1$). The full line is plotted with $\beta= 0.88$ and 
experimental parameters of Ref.\protect \cite{c16} and shows that the position 
is unsqueezed ($\frac{\Delta}{\Omega_m} = 1$; $\langle\delta 
x_m^2\rangle\approx4.44$). The cercled line is plotted for $\beta =0.1$ and shows 
that the position is squeezed ($\frac{\Delta}{\Omega_m} = 1$; $\langle\delta 
x_m^2\rangle\approx0.69$).}
\label{fig:Fig2}
\end{figure}

The dynamics of mechanical fluctuations is obtained by writing Eqs. 
(\ref{eq:4}) in the Fourier space,
\begin{subequations}
\label{eq:5}
\begin{equation}
 B(\Omega)u(\Omega)+n(\Omega)=0,
\end{equation}
where
\begin{equation}
 B(\Omega)=\begin{pmatrix}i\Omega & \Omega_m & 0 &0\\\Omega_m(\beta-1) &
(i\Omega -\Gamma_m) & G & 0\\0 & 0 & (i\Omega-\frac{\kappa}{2}) &
-\tilde{\Delta}\\ G & 0 & \tilde{\Delta} & (i\Omega-\frac{\kappa}{2})
\end{pmatrix},
\end{equation}
\end{subequations}
Solving the matrix equation straightforwardly, we obtain the solution for the
mechanical displacement operator to be
\begin{equation}
\begin{split}
\chi_{eff}^{-1}(\Omega)\delta x_m(\Omega) & =a_1G\Omega_m\sqrt{\kappa}
\left(\tilde{\Delta}^2+\frac{\kappa}{4}^2-\omega^2+i\kappa\Omega\right)
\\
& \times\left[-\tilde{\Delta}\delta\varphi^{in}+\left(-i\Omega+\frac{\kappa}{2}
\right)\delta I^{in} \right]+\Omega_m F_{th},
\end{split}
\end{equation}
where
\begin{equation}
 a_1=\left[ \left(\tilde{\Delta}^2+\frac{\kappa}{4}^2-\Omega^2\right)^2
+\kappa^2\Omega^2\right] ^{-1},
\end{equation}
and
\begin{equation}
\chi_{eff}(\Omega)=(\Omega_{eff}^2-\Omega^2-i\Omega\Gamma_{eff})^{-1},
\label{eq:u}
\end{equation}
is the effective susceptibility of the oscillator with the effective resonance
frequency and damping rate given by
\begin{equation}
\Omega_{eff}^2(\Omega)=\Omega_m^2\left(1+a_1G^2\frac{\tilde{\Delta}}{\Omega_m}
\left(\tilde{\Delta}^2+\frac{\kappa}{4}^2-\Omega^2\right)-\beta \right),
\label{eq:m}
\end{equation}
\begin{equation}
\Gamma_{eff}(\Omega)=\Gamma_m-a_1G^2\Omega_m\tilde{\Delta}\kappa.
\end{equation}
By using the correlation functions of the noise sources for a coherent beam
in the frequency domain, the oscillator position and the momentum variances are
defined by,
\begin{align}
\langle\delta x_m^2\rangle & =\frac{1}{2\pi}\int_{-\infty}^{+\infty}d\Omega
|\chi _{eff}|^2S_x,  \label{eq:f} \\
\langle\delta p_m^2\rangle & =\frac{1}{2\pi}\int_{-\infty}^{+\infty}d\Omega
\frac{\Omega^2}{\Omega_m^2}|\chi_{eff}|^2S_x,
\label{eq:i}
\end{align}
where the position noise spectrum is given by
\begin{equation}
\begin{split}
S_x(\Omega)& =a_1G^2\Omega_m^2\frac{\kappa}{2}\left(\tilde{\Delta}^2-\Omega^2
+\frac{\kappa}{4}^2-i\kappa\tilde{\Delta}\right)\\
& +2\Gamma_m\Omega_m\Omega\left(1+\coth\left(\frac{\hbar\Omega}{2k_{B}T}
\right)\right).
\end{split}
\label{eq:t}
\end{equation}%
At the quasi resonant frequency ($\Omega\approx \Omega_m$) and for $\coth\left(
\frac{\hbar\Omega}{2k_{B}T}\right)\approx\frac{2k_{B}T}{\hbar\Omega}$ (satisfied
with experimental parameters used), the exact solutions of integrals
(\ref{eq:f}) and (\ref{eq:i}) are given by
\begin{align}
\langle\delta x_m^2\rangle & =\frac{\Omega_m^2}{4\Gamma_{eff}\Omega_{eff}^2}a_2,
\label{eq:j}\\
\langle \delta p_m^2\rangle & =\frac{1}{4\Gamma_{eff}}a_2,
\label{eq:k}
\end{align}
where
\begin{equation}
 a_2=\frac{\frac{G^2}{\Omega_m^2}\kappa \left(
\frac{\tilde{\Delta}^2}{\Omega_m^2}+\frac{\kappa^2}{4\Omega_m^2}-1-i\frac{\kappa
\tilde{\Delta}}{\Omega_m^2}\right)}{\left(\frac{\tilde{\Delta}^2} {\Omega_m^2}
+\frac{\kappa^2}{4\Omega_m^2}-1\right)^2+\frac{\kappa^2}{\Omega_m^2}}
+4\Gamma_m\left( 1+\frac{2k_{B}T}{\hbar \Omega_m}\right).
\end{equation}
These position and momentum variances should satisfy the Heisenberg relation,
\begin{equation}
\langle\delta x_m^2\rangle\langle\delta p_m^2\rangle\geq|\frac{1}{2}[x_m,p_m]
|^2,
\end{equation}
 that is,
\begin{equation}
\langle\delta x_m^2\rangle\langle\delta p_m^2\rangle\geq1.
\label{eq:l}
\end{equation}%
There is no condition on the individual quadratures of relation (\ref{eq:l}).
However, for a standard quantum limit (SQL) of $1$, the coherent states must
satisfy $\langle\delta x_m^2\rangle=\langle \delta p_m^2\rangle=1$. When one
variance is below the SQL, i.e., $\langle\delta x_m^2\rangle<1$ or $\langle
\delta p_m^2\rangle<1$, the corresponding quadrature is said to be squeezed.
Generally, the squeezed states are characterized by the asymmetry between its
quadratures which is mostly introduced by the nonlinear effects. According to
Eqs. (\ref{eq:m}) and (\ref{eq:j}), only $\langle\delta x_m^2\rangle$ depends on
the geometrical nonlinearity through the term $\frac{\Omega_m^2}{\Omega_{eff}^2
}$. Contrariwise, the value of $\langle \delta p_m^2\rangle$ is obtained by
using experimental parameters of Ref. \cite{c16} in Eq. (\ref{eq:k}) and it
appears to be squeezed up to about $37\%$ (see Fig.\ref{fig:Fig1}). Therefore,
the position variance is deduced from the mean energy of the nanoresonator in
the steady state,
\begin{equation}
E=\frac{\hbar\Omega_m}{4}(\langle\delta x_m^2\rangle+\langle\delta p_m^2\rangle)
\equiv\hbar\Omega_m\left(n_{eff}+\frac{1}{2}\right),
\label{eq:x}
\end{equation}
by substituting the effective phonon number with the experimental value
$n_{eff}=0.85\pm 0.08$ of Ref. \cite{c16}. The value obtained is $\langle
\delta x_m^2\rangle \approx 4.44$ which is unsqueezed (see Fig.
\ref{fig:Fig2}).

\begin{figure}[htbp]
\centering
\begin{minipage}{16cm}
 \begin{center}
\resizebox{0.80\textwidth}{!}{
\includegraphics{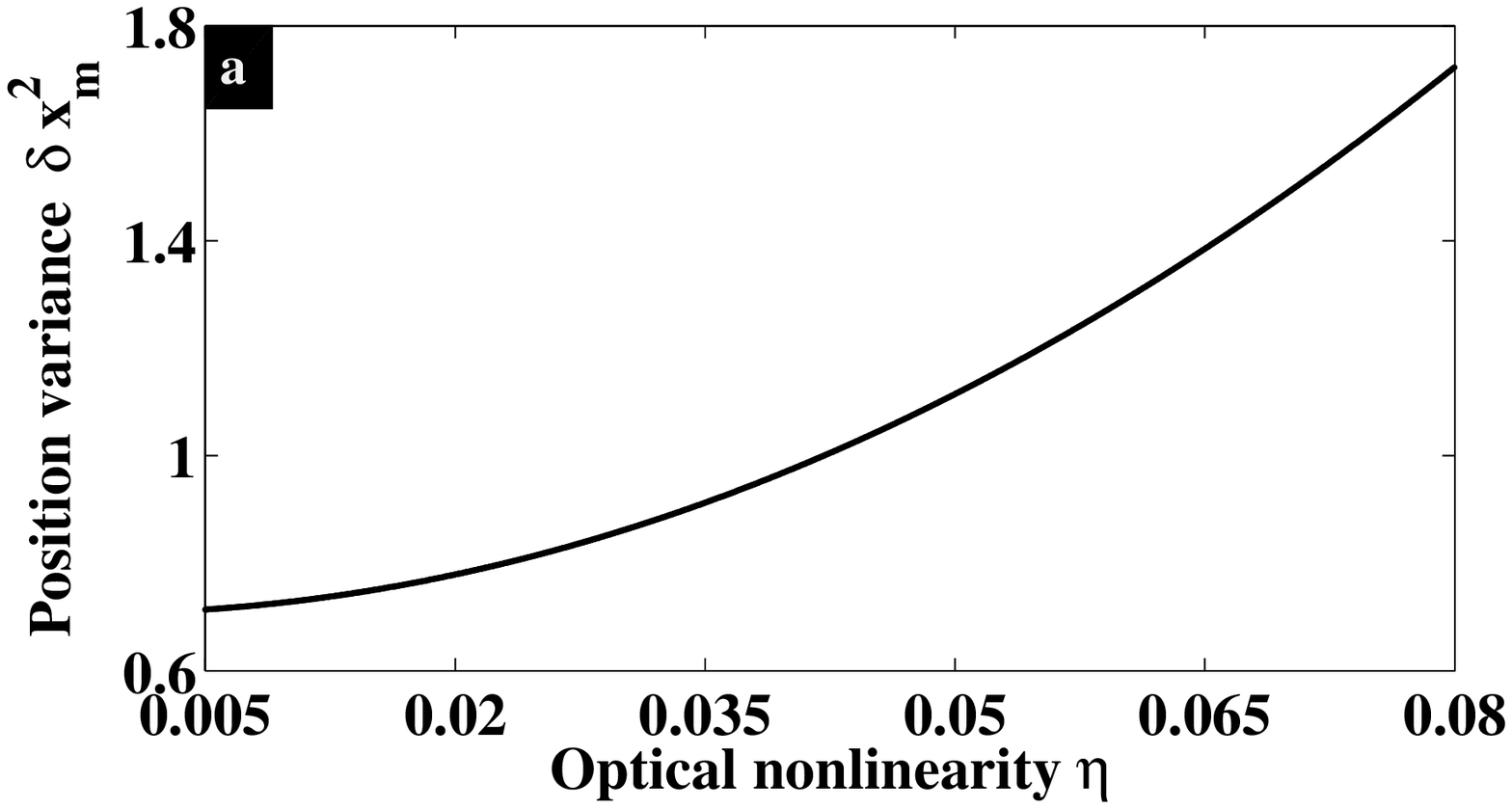}}
\resizebox{0.80\textwidth}{!}{
\includegraphics{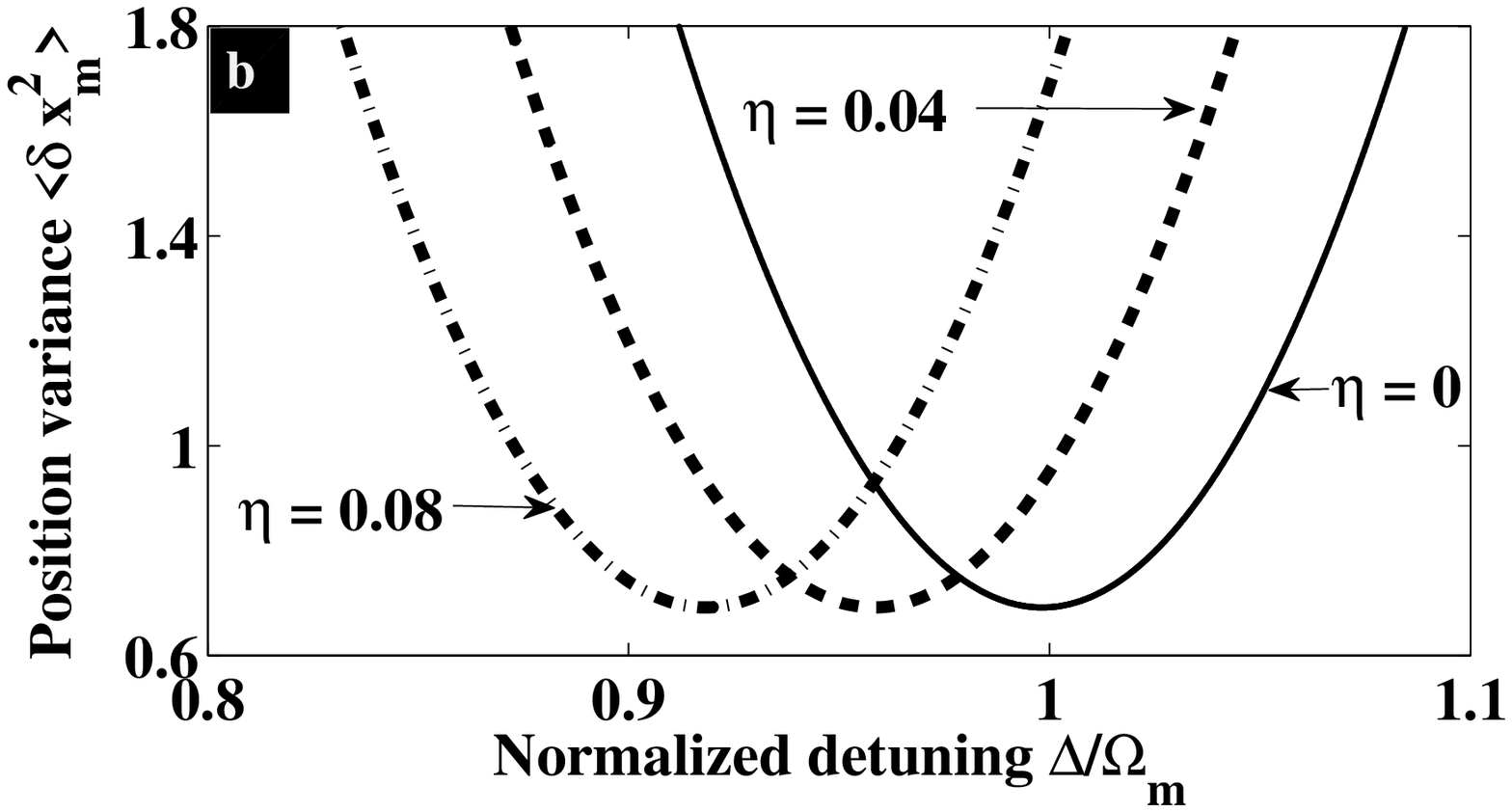}}
\resizebox{0.80\textwidth}{!}{
\includegraphics{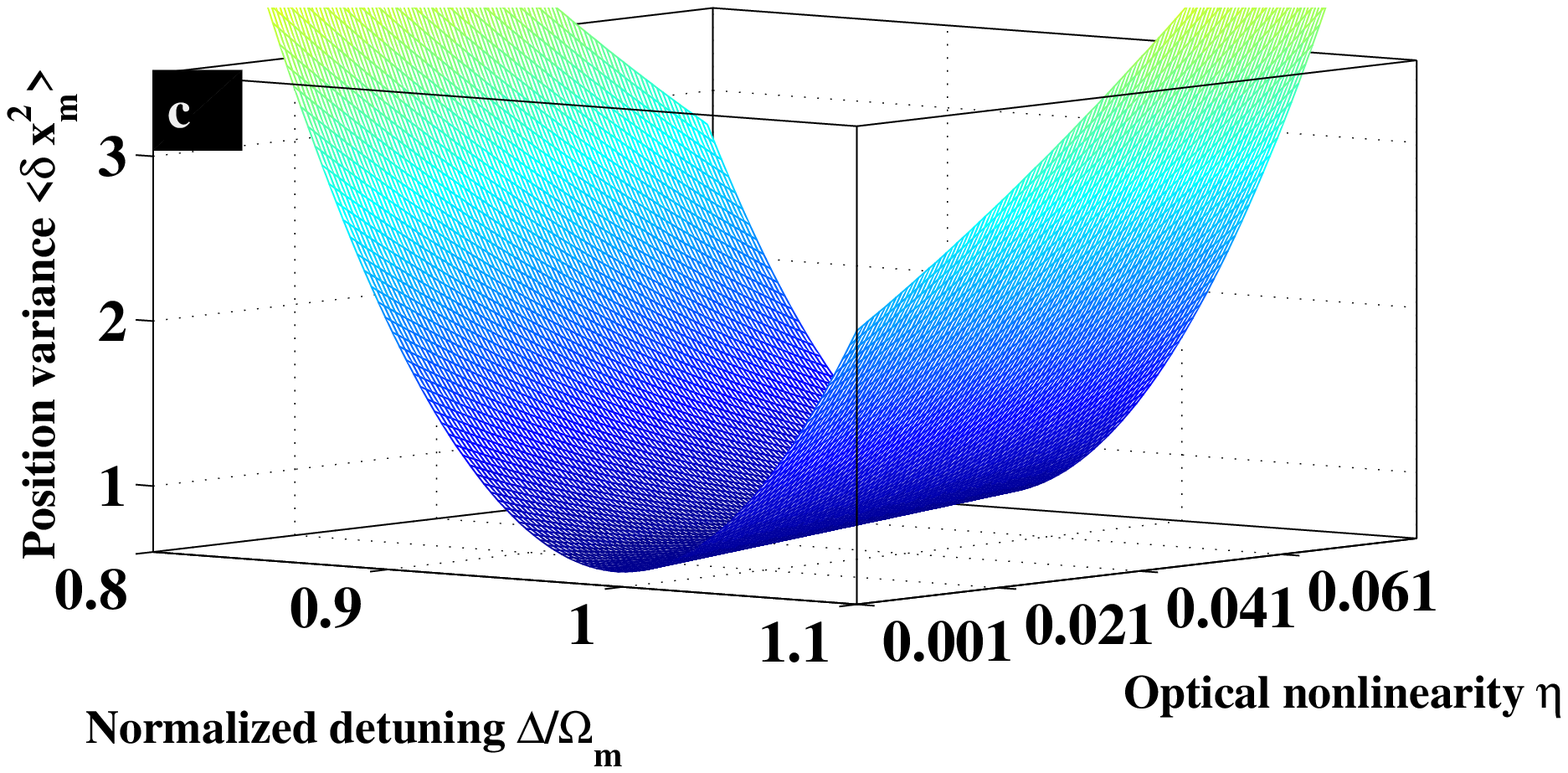}}
\caption{Effect of optical nonlinearity $\eta$ on the position variance for
$\beta = 0.1$. a. Position variance versus $\eta$ shows that $\langle\delta 
x_m^2\rangle$ becomes unsqueezed when $\eta$ increases. b. shows that optimum 
squeezing shifts towards the left when $\eta$ increases. c. shows the combined effects of 
$\eta $ on $\langle\delta x_m^2\rangle$.}
\label{fig:Fig3} 
\end{center}
\end{minipage}     
\end{figure}

Indeed, the ratio $\frac{\Omega_m^2}{\Omega _{eff}^2}$ increases when $\beta$
increases (for the high mechanical displacements) and rises up the position
variance $\langle \delta x_m^2\rangle$. So, as the re-thermalization time of the
mechanical resonator or the decoherence time \cite{c17,c16}, the geometrical
nonlinearity limits the squeezing and some quantum effects \cite{c151}. In fact,
$\beta $ depends on the bending moment of the resonator and takes into account
its internal vibrations. These internal vibrations increase for large bending
and contribute to the re-thermalization of the resonator at the low
temperatures. These effects are reversed to those of the Kerr nonlinearities
which improve the squeezing \cite{c8}. For $\langle\delta x_m^2\rangle \approx
4.44$, $\beta$ is evaluated to be about $0.88$ (see Fig. \ref{fig:Fig2}) and is
in the domain given in Table \ref{tab:aa} and corresponds to the high mechanical
displacements for the nanoresonators \cite{c18}. It also appears that for
$\beta=0.1$, the position variance $\langle \delta x_m^2\rangle$ is under the
standard quantum limit (see Fig. \ref{fig:Fig2}), allowing therefore the
squeezing (up to about $31\%$).
In order to investigate the effect of the optical nonlinearity $\eta$ on the 
mechanical squeezing, we consider $\eta \neq 0$ in the expression 
$\frac{\tilde{\Delta}}{\Omega_m} = \frac{\Delta}{\Omega_m} + \eta $ which 
appears in Eqs. (\ref{eq:j}) and (\ref{eq:k}). Figs. \ref{fig:Fig3} shows the 
effect of $\eta$ on the position variance $\langle\delta x_m^2\rangle$ for 
$\beta = 0.1$. One remarks in Fig. \ref{fig:Fig3}.a that the position variance 
$\langle\delta x_m^2\rangle$ becomes unsqueezed for high values of $\eta$ 
($\eta>0.042$). This effect of $\eta$ is similar to that of the $\beta$ shown 
in Fig. \ref{fig:Fig2}. It appears in Fig. \ref{fig:Fig3}.b that  $\eta$ shifts 
the optimal squeezing towards the left. Since $\eta$ is always present in 
optomechanical systems, it is then important to quantify it in experiments in 
order to determine exactly at which detuning the optimal squeezing can be 
evaluated. On the Fig. \ref{fig:Fig3}.c where the two mentioned effects of 
$\eta$ are represented, $\langle\delta x_m^2\rangle$ increases with $\eta$ 
and the optimal position squeezing is not always at $\frac{\Delta}{\Omega_m}= 
1$ but depends on the value of $\eta$ in the range $\frac{\Delta}{\Omega_m} 
\in[0.9;1]$. One also notes that the effects of $\eta$ on the momentum variance 
are the same as these described on Figs. \ref{fig:Fig3}.  

On the other hand, the squeezing is improved despite such nonlinearities in the 
nanobeams when the temperature is tuned down and/or the system is carried in a 
regime of strong optomechanical coupling.

\section{Squeezing of the optical output quadratures}\label{sec:SqOpt}

From the matrix equation (\ref{eq:5}) we also obtain the solutions for the
intracavity phase and intensity quadratures operators to be
\begin{equation}
\delta I =-a_3\left[ \tilde{\Delta}G\delta x_m-\tilde{\Delta}\sqrt{\kappa}
\delta\varphi^{in}+\sqrt{\kappa }\left(-i\Omega+\frac{\kappa}{2}\right)\delta
I^{in}\right] ,\label{eq:n}
\end{equation}
and
\begin{equation}
\delta\varphi =a_3\left[(G\delta x_m+\sqrt{\kappa}\delta\varphi^{in})
\left(-i\Omega+\frac{\kappa}{2}\right)+\tilde{\Delta}\sqrt{ \kappa} \delta
I^{in}\right] ,\label{eq:o}\\
\end{equation}
with
\begin{equation}
a_3  =\left[ \left( -i\Omega +\frac{\kappa }{2}\right)^2+\tilde{\Delta}^2
\right] ^{-1}.
\end{equation}
In order to analyze their squeezing, we use the well-known input-output relation
\cite{c241}
\begin{equation}
\alpha^{out}=-\alpha^{in}+\sqrt{\kappa}\alpha,\label{eq:p1}
\end{equation}
and then deduce
\begin{equation}
\delta I^{out}  =-a_3\left[ \tilde{\Delta}\sqrt{\kappa}G\delta x_m-\tilde\Delta
\kappa\delta\varphi^{in}+\left(\Omega^2+\frac{\kappa^2}{4}-\tilde{\Delta}
^2\right)\delta I^{in}\right] , \label{eq:p}
\end{equation}
and
\begin{equation}
\delta\varphi^{out} =a_3\left[ \sqrt{\kappa}G\left(-i\Omega+\frac{\kappa}{2}
\right)\delta x_m+\left(\Omega^2+\frac{\kappa^2}{4}-\tilde{\Delta}^2\right)
\delta\varphi^{in}+\tilde{\Delta}\kappa\delta I^{in}\right] .
\label{eq:q}
\end{equation}
By using the spectral density $S_{A^{out}}(\Omega) =\frac{1}{2\pi}\int_{-\infty
}^{+\infty}d\omega e^{-i(\Omega+\omega)t}\langle \delta A^{out}(\Omega)\delta
A^{out}(\omega)\rangle$, we obtain the output spectrum of the intensity and
phase as,
\begin{align}
\begin{split}
S_{I^{out}}(\Omega) & =a_1\tilde{\Delta}^2G^2\kappa|\chi_{eff}|^2S_x(\Omega)
+\frac{a_1}{2}\left(\Omega^2+\frac{\kappa^2}{4}-\tilde{\Delta}^2\right)^2\\
& +a_1\tilde{\Delta}^2\kappa^2+A(\Omega)-B(\Omega)-C(\Omega),
\label{eq:r}
\end{split}\\
\begin{split}
S_{\varphi^{out}}(\Omega)& =a_1G^2\kappa\left(\frac{\kappa^2}{4}
+\Omega^2\right)|\chi_{eff}|^2S_x(\Omega) +\frac{a_1}{2} 
\left(\Omega^2+\frac{\kappa^2}{4}-\tilde{\Delta}^2\right)^2\\
& +a_1\tilde{\Delta}^2\kappa^2+G^2\kappa\Omega_m(D(\Omega)+E(\Omega)),
\label{eq:s}
\end{split}
\end{align}
where,
\begin{align}
A(\Omega) & =a\tilde{\Delta}G^2\Omega_m\kappa(\Omega_{eff}^2-\Omega^2)
(\tilde{\Delta}-\Omega), \\
B(\Omega) & =b\tilde{\Delta}G^2\Omega_m\kappa\Omega\Gamma_{eff}
(\tilde{\Delta}-\Omega), \\
C(\Omega) & =\tilde{\Delta}G^2\frac{\kappa^2}{4}\Omega_m[2b(\Omega_{eff}^2
-\Omega^2)-2a\Omega\Gamma_{eff}],\\
D(\Omega) & = \left[ (\Omega_{eff}^2-\Omega^2)\left(\tilde{\Delta}^2
+\frac{\kappa^2}{4}-\Omega^2\right)+\kappa\Omega^2\Gamma_{eff} \right]
 \left(\frac{c}{2}-\Omega d\right)\kappa a_1,\\
E(\Omega) & =\left[ \kappa(\Omega_{eff}^2-\Omega^2)-\Gamma_{eff}
\left(\tilde{\Delta}^2 +\frac{\kappa^2}{4}-\Omega^2\right)\right]
\left(\Omega c+\frac{\kappa^2}{2}d\right)\Omega a_1,
\end{align}
with
\begin{align}
a &=a_1^2\kappa\left[ \tilde{\Delta}\left(\tilde{\Delta}^2+\frac{\kappa^2}{4}
-\Omega^2\right)-\Omega\left(\Omega^2+\frac{\kappa^2}{4}-\tilde{\Delta}^2
\right)\right] ,\\
b &=a_1^2\left[ \left(\tilde{\Delta}^2+\frac{\kappa^2}{4}-\Omega^2\right)
\left(\Omega^2+\frac{\kappa^2}{4}-\tilde{\Delta}^2\right) +\tilde{\Delta}
\kappa^2\Omega\right] ,\\
c & =2(\Omega-\tilde{\Delta})\left(\Omega^2+\frac{\kappa^2}{4}
-\tilde{\Delta}^2\right)+\kappa^2\tilde{\Delta} \\
d & =2\tilde{\Delta}(\Omega-\tilde{\Delta})-\left(\Omega^2+\frac{\kappa^2}{4}
-\tilde{\Delta}^2\right).
\end{align}
In Eqs. (\ref{eq:r}) and (\ref{eq:s}), the first terms are proportional to the
position spectrum $S_x$ (Eq. (\ref{eq:t})) and to the effective mechanical
susceptibility (Eq. (\ref{eq:u})). These terms derive from the mechanical
fluctuations of the oscillator. The other terms in (\ref{eq:r}) and (\ref{eq:s})
originate to the fluctuations of the input beam. Assuming that the system is in
the quasi resonant regime ($\Omega\approx\Omega_m$), all the contributions
related to the input beam fluctuations take constant values. The expressions
(\ref{eq:r}) and (\ref{eq:s}) can now be integrated. By using the residues
theorem and the Cauchy-Goursat theorem, one readily obtains
\begin{equation}
\langle\delta I^{out\,2}\rangle =\frac{\frac{\tilde{\Delta}^2}{\Omega_m^2}
\frac{G^2}{\Omega_m^2}\kappa}{\left(\frac{\tilde{\Delta}^2}{\Omega_m^2}
+\frac{\kappa ^2}{4\Omega_m^2}-1\right)^2+\frac{\kappa^2}{\Omega_m^2}}
\langle\delta x_m^2\rangle,\label{eq:v}
\end{equation}
and
\begin{equation}
\langle\delta\varphi^{out\,2}\rangle =\frac{\left(1+\frac{\kappa^2}
{4\Omega_m^2}\right)\frac{G^2}{\Omega_m^2}\kappa}{\left(\frac{\tilde{\Delta}^2}
{\Omega_m^2}+\frac{\kappa^2}{4\Omega a_m^2}-1\right)^2+\frac{\kappa^2}
{\Omega_m^2}}\langle\delta x_m^2\rangle,\label{eq:w}
\end{equation}
where $\langle\delta x_m^2\rangle$ is given by Eq. (\ref{eq:j}).

At the mechanical resonance ($\tilde{\Delta}\approx {\Omega_m}$) where it is
established above that the position variance is unsqueezed ($\langle\delta x_m^2
\rangle=4.44$), we deduce from Eqs. (\ref{eq:v}) and (\ref{eq:w}) that both
optical variances are unsqueezed. However, the effective detuning
$\tilde{\Delta} \approx\Delta+g_M\bar{x}_m$ leads at the optical resonance
(${\Delta=0}$) to
\begin{equation}
\frac{\tilde{\Delta}}{\Omega_m}\approx \frac{g_M\bar{x}_m}{\Omega_m}=\eta.
\label{eq:z}
\end{equation}

\begin{figure}[htpb]
\centering
\includegraphics[scale=.70]{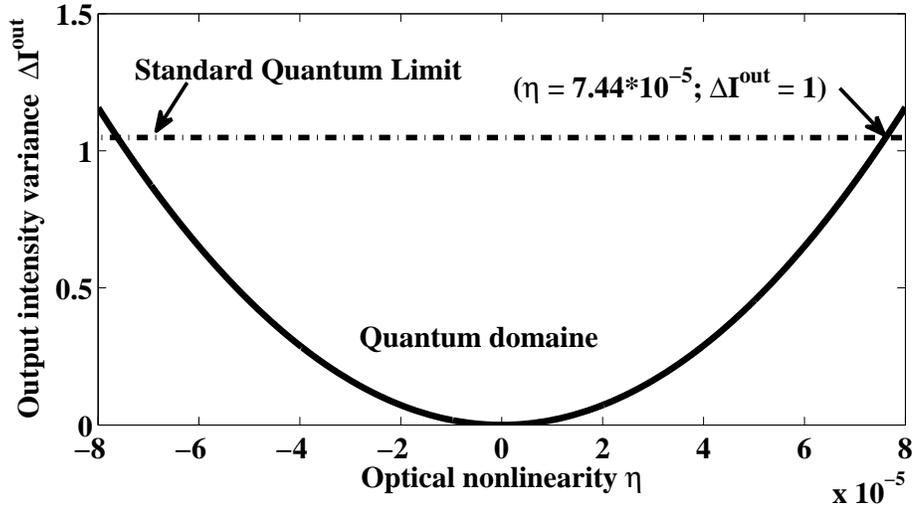}
\caption{Plot of the mean square fluctuation of the output light $\Delta 
I^{out}$ versus the optical nonlinearity $\eta $ for $\beta=5.72\times10^{-4}$. 
With the low input power used for the system $P_{in}\leq 30\operatorname{\mu 
W}$ \protect \cite{c17}, the output intensity is squeezed ($|\eta|\leq 
7.44\times10^{-5}$).}
\label{fig:Fig4}
\end{figure}

\begin{figure}[htb]
\centering
\begin{minipage}{16cm}
 \begin{center}
\resizebox{0.80\textwidth}{!}{
\includegraphics{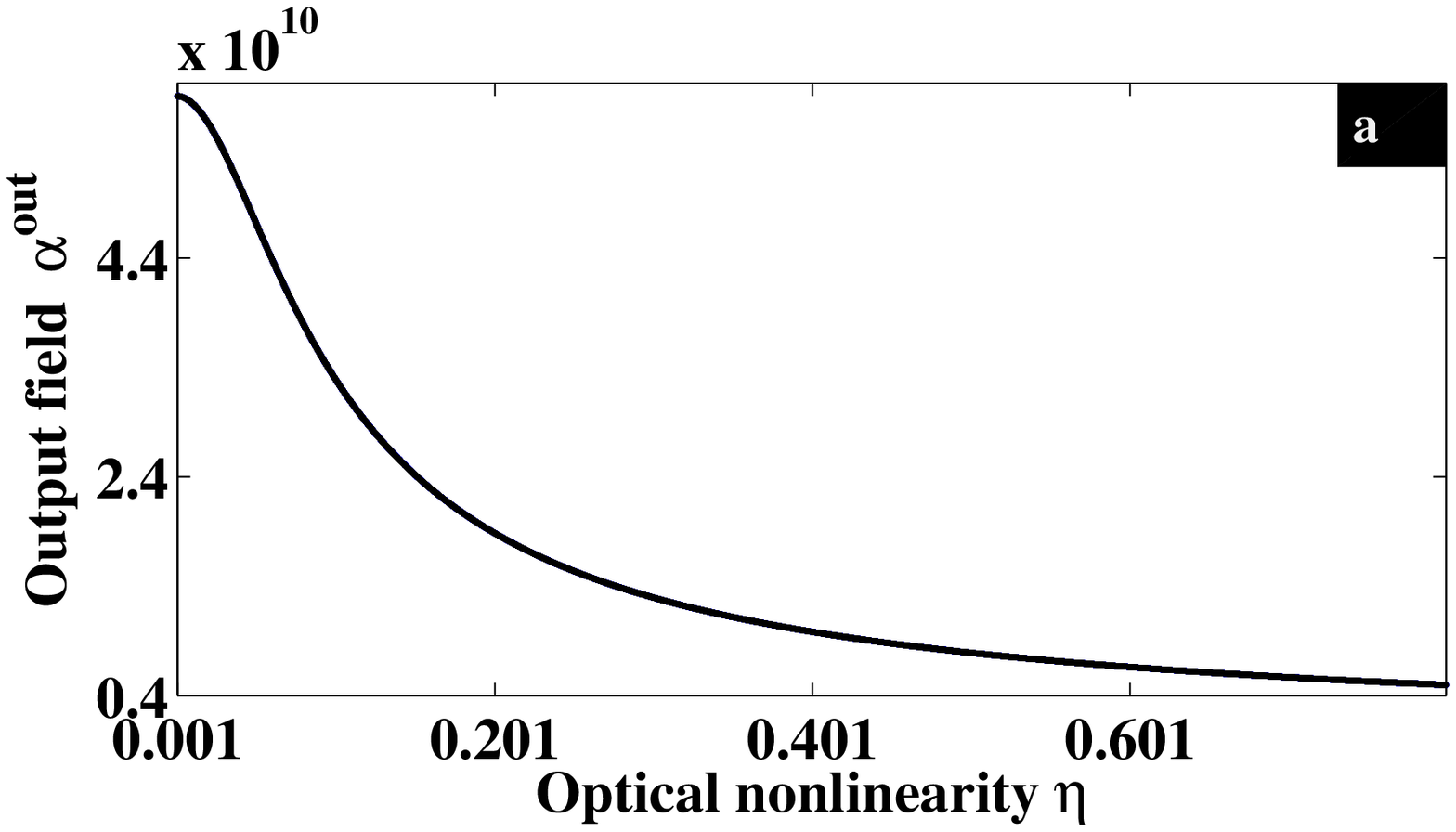}}
\resizebox{0.80\textwidth}{!}{
\includegraphics{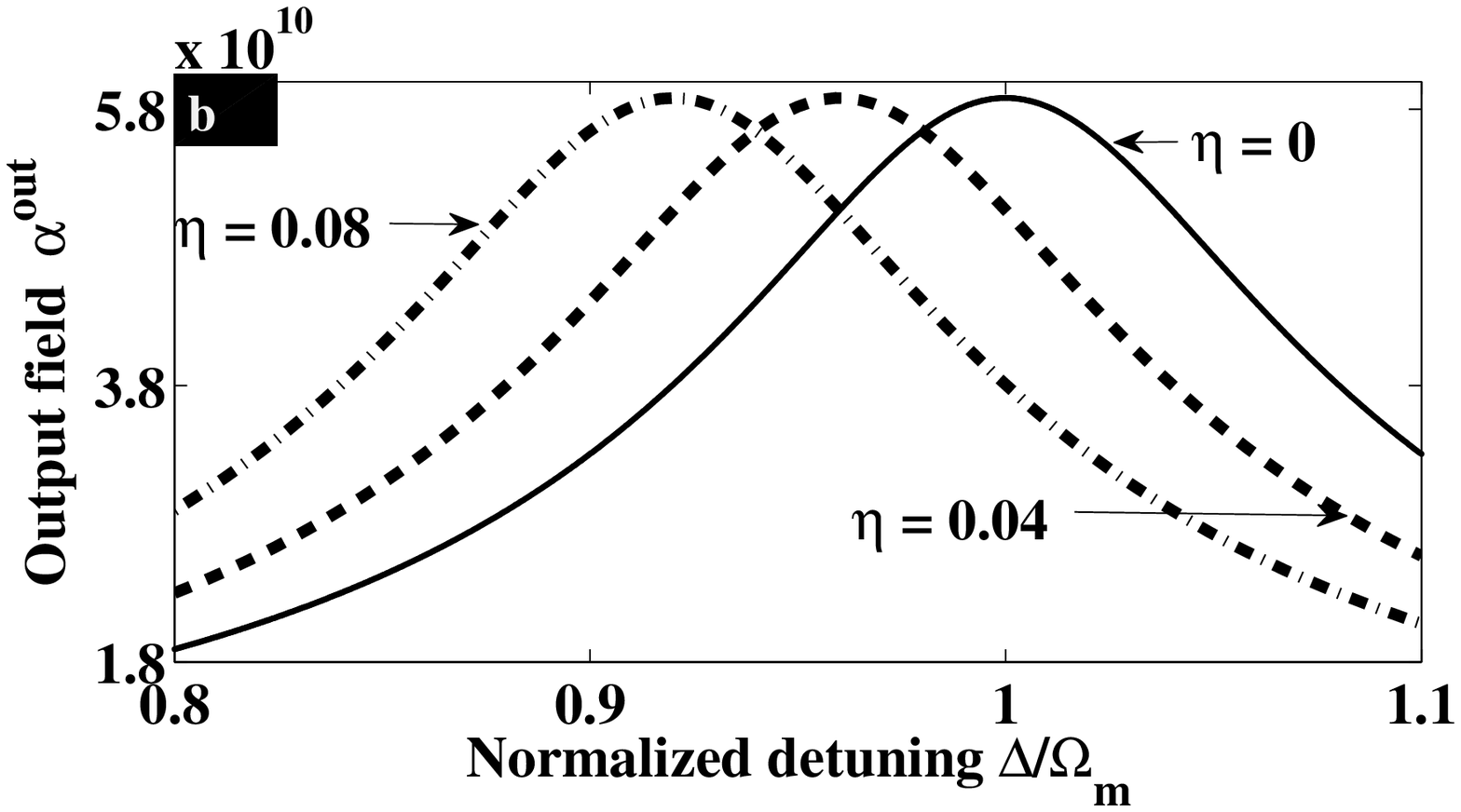}}
\resizebox{0.80\textwidth}{!}{
\includegraphics{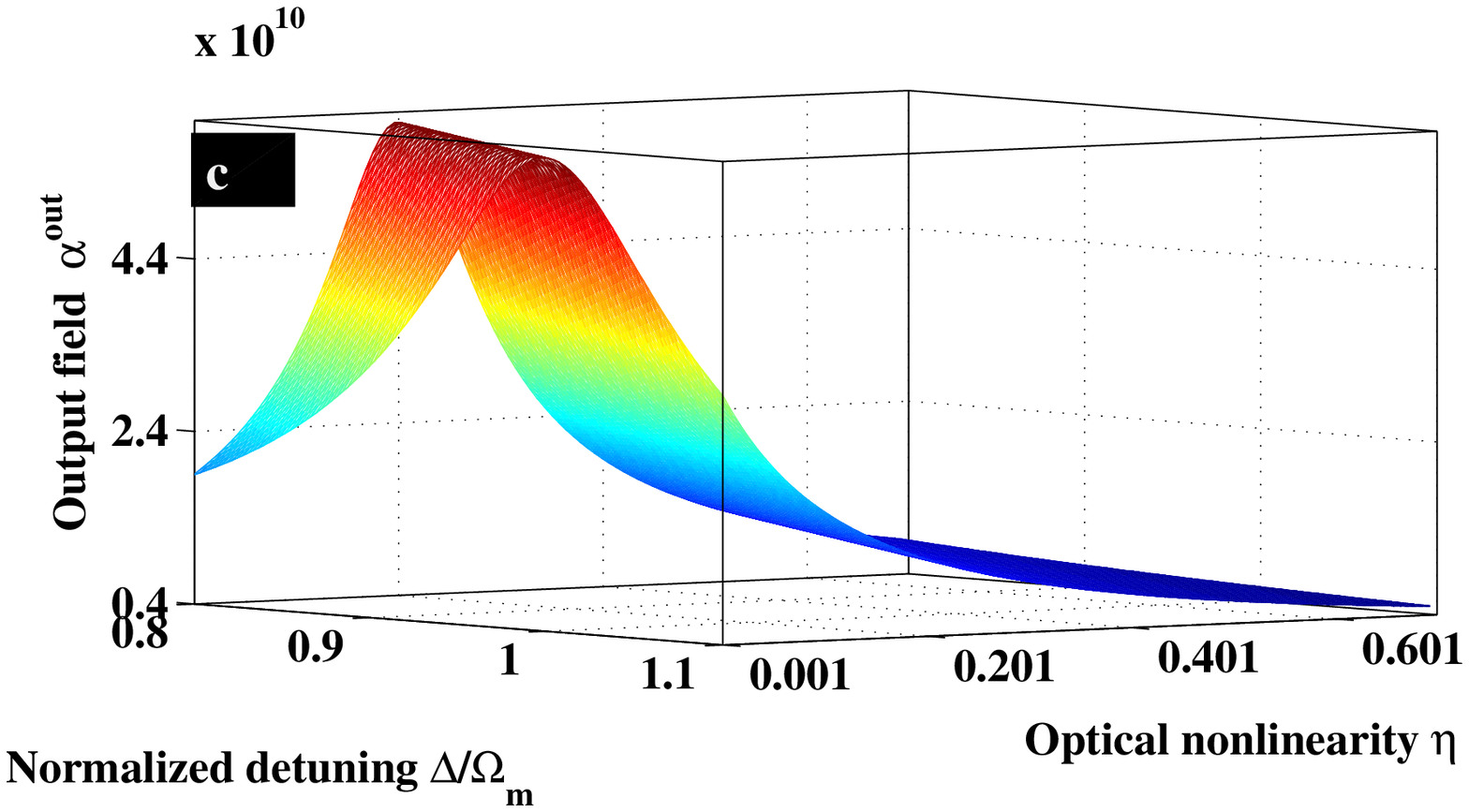}}
\caption{Effect of optical nonlinearity $\eta$ on the output field. a. Output 
field versus $\eta$ shows that $\alpha ^{out}$ decreases when $\eta$ increases. 
b. The optimal output field shifts towards the left when $\eta$ increases. 
c. Combined effects of $\eta$ on $\alpha ^{out}$.}
\label{fig:Fig5} 
\end{center}
\end{minipage}     
\end{figure}

Fig. \ref{fig:Fig4} shows that $\Delta I^{out}$ is squeezed when $|\eta|\leq
7.44\times 10^{-5}$ which is obtained with the input power $P_{in}\leq
30\operatorname{\mu W}$ \cite{c17}. This means that at the optical resonance and
for the low input power, the optomechanical cavity behaves as an optical filter
or \emph{noise eater} \cite{c10}. Thus, the quantum fluctuations (shot noise) in
the input coherent laser beam ($\Delta I^{in}=1$) are reduced after been
reflected out of the optomechanical cavity ($\Delta I^{out}<1$). To improve this
squeezing at the optical resonance one can increase $\eta$ (for the negative
values of the detuning) or decrease $\eta$ (for the positive values of the
detuning) (see Fig. \ref{fig:Fig4}).
By using Eqs.(\ref{eq:a1}) and (\ref{eq:p1}), one can expressed output field as
\begin{equation}
\alpha ^{out}=\frac{a_4+ia_5}{\left( \frac{\Delta }{\Omega_m}+\eta -1\right)
^2+\frac{\kappa^2}{4\Omega_m^2}}
\end{equation}
with
\begin{align}
a_4&=\frac{\kappa^2}{4\Omega_m^2}\alpha ^{in}+\left( \frac{\Delta }{%
\Omega_m}+\eta -1\right) \left( \left( \frac{\Delta }{\Omega_m}+\eta
-1\right) \alpha ^{in}+\frac{\sqrt{\kappa }}{\Omega_m}\varepsilon
^{in}\right),\\ 
a_5&=\left( \left( \frac{\Delta }{\Omega_m}+\eta -1\right) 
\frac{\kappa }{\Omega_m}\alpha ^{in}-\frac{\kappa \sqrt{\kappa }}{2\Omega
_{m}^2}\varepsilon ^{in}\right), 
\end{align}
where $\varepsilon^{in}=\sqrt{\frac{2\kappa P_{in}}{\hbar \Omega _m}}$. One
remarks that $\alpha^{out}$ depends only on the optical nonlinearity which
allows us to quantify his effect on the output field. Fig. \ref{fig:Fig5}.a
shows that the output field decrease when $\eta$ increases. This means that it
is important to control the value of $\eta$ in order to obtain the output field
needed. This control of $\eta$ also gives the value of the detuning at which the
output field is optimal (see Fig. \ref{fig:Fig5}.b). Fig. \ref{fig:Fig5}.c shows
the two mentioned effects of $\eta$ on the output field $\alpha ^{out}$. 

It is also found that, as for the position, the decrease of temperature
induces an improvement of intensity squeezing. Regarding the geometrical
nonlinearity, it contributes to reduce the squeezing when it becomes large.
But, it is generally weak around the optical resonance (see Table \ref{tab:aa}),
so its effects are neglected at this sideband.

Let us recall that the reduction of the nonlinearities in the optomechanical
systems, constitutes the key of many future quantum optomechanical applications.
This requires the reduction of the mean displacement $\bar{x}$ of the
nanoresonator or the increase of the fundamental mechanical frequency
$\Omega_m$. This consists to use a very high finesse cavity which can be excited
by the low input power \cite {c17}.

It should be noted that squeezing of nonlinear optomechanical systems in which 
an optical cavity mode is coupled quadratically rather than linearly to the 
position of mechanical oscillator have been studied in Ref. \cite{c24}. While 
Sete and Eleuch, in Ref. \cite{c152}, investigating nonlinear effects in an 
optomechanical system containing a quantum well, have found that as a result of 
the nonlinearity induced by the optomechanical coupling, the transmitted field 
exhibits strong squeezing at certain hybrid resonance frequencies and system 
parameters.

\section{Conclusion}\label{sec:Concl}

We have presented an analytical study of the geometrical and optical nonlinear
effects on the optomechanical squeezing. Contrary to the Kerr nonlinearities, it
is found that these two nonlinearities reduced the squeezing. At the detuning
$\tilde{\Delta}\approx\Omega_m$ where the displacement is important, the
momentum is squeezed while the position squeezing is very restricted for the
small values of $\beta$. This effect is justified by the geometrical
nonlinearity which depends on the bending moment of the resonator and takes into
account its internal vibrations. At the optical resonance, the output intensity
is squeezed when $|\eta|$ is small ($|\eta|<7.44\times 10^{-5}$).

In a future work, we will investigate the squeezing as a function of the 
scanning frequency ($\Omega_m$) in order to study the squeezing at different 
resonances. 

% \section*{Acknowledgments}
% % \begin{acknowledgments}
%  The authors would like to express their gratitude to the referees, whose 
% valuable comments have improved the paper.
% \end{acknowledgments}

\end{document}